\documentstyle[epsf]{article}
\begin{document}
\large
\baselineskip 10mm
\begin{center}
{\huge Species Orthogonalization}
\end{center}

\baselineskip 9mm
\baselineskip 6mm

\vspace{1mm}
\begin{center}
{\Large Petr Kr\'al}
\end{center}

\begin{center}
{\it Department of Chemical Physics, 
      Weizmann Institute of Science,\\
         76100 Rehovot, Israel}
\end{center}

\begin{center}
{\it Present address:
ITAMP, Harvard-Smithsonian Center for Astrophysics,
         Cambridge, Massachusetts 02138}
\end{center}

\vspace{10mm}
{\noindent
We discuss general formation of complementary behaviors, functions and forms 
in biological species competing for resources.  We call {\it orthogonalization} 
the related processes on macro and micro-level of a self-organized formation 
of correlations in the species properties. Orthogonalization processes 
could be, for example, easily observed in sympatric speciation, as we show 
in numerical studies carried with a new population equation.  As a practical 
result, we find that the number of species is proportional to the effective 
richness of resources and depends on their history.}
\vspace{10mm}

\begin{center}
{\Large Introduction}
\end{center}
Individuals in any biological species differ a little in their behaviors, 
forms and other parameters, which can help to relax their otherwise large 
competition. If the differences become too large, individuals can decrease 
their interest for interbreeding, and species can consequently split up. During
speciation and species coexistence correlations are formed between parameters 
of the species, which adjust distributions for different species in such 
a way that resources are effectively shared. 
In this process complex patterns of complementary sets of parameters are 
formed, where species can be seen to approach discrete {\it orthogonal 
states}. These ``orthogonalization processes" (OP) can be modeled in 
multidimensional mathematical spaces, with coordinates given by the 
species parameters. 

It appears natural to define species by a mutual reproducibility of their 
individuals (Mayr (1942, 1963)). Since OP are strong in clusters of 
evolutionary and functionally close species, they could give a clear 
substance to the species definitions.  Globally, we can define {\it species 
orthogonalization} as a process in which species approach a state, where 
they share resources 
and habitats in the most effective way, fully stabilized in time.  Since 
species not only passively approach this ideal state, but they prefer 
to overturn the ``status quo", by permanently improving the quality of 
their gene equipment, full orthogonalization can not be realized.  Instead, 
evolution gives a relatively stable coexistence (quasi-orthogonality) 
of biological species, time from time disturbed by speciation events. 
Close to these events OP are the most intense, but they are strong until
the new species ``settle down". Species interactions, which enable this
process, last even after the species get in the resulting orthogonal states.

The need for orthogonalization might have caused the disappearance of 
many ``intermediate" species, since these would hamper orthogonalization
in the short and long term run, the last because {\it both} the resources 
volume and speciation rate are relatively stable. A more specific trace 
of OP are {\it character shifts} observed in species sharing habitats 
or in sympatric speciation (Brown \& Wilson (1956), Smith (1966), 
May \& Mac Arthur (1972), Slatkin (1980), Schluter \& Mac Phail  (1988), 
Doebeli (1996a), Drossel \& Mac Kane (1999)), which reveal efficient
division of resources from the point of structure, day time or form of 
consumption.  Competition for (orthogonal) discrete niches (Hutchinson 
(1968)) is confirmed, for example, by simple speciation/extinction models 
(Valentine \& Walker (1988)). Therefore, OP might even control the 
numbers of species living in given areas (Rosenzweig (1995), 
Plotkin {\it et al.} (2000)).  

To understand OP, we discuss first some evolution aspects of species on 
the micro and macro-level, and flow of information between the levels. 
Then, we focus on modeling of a sympatric speciation of species (Doebeli 
(1996b), Bagnoli \& Bezzi (1997), Kondrashov \& Kondrashov (1999), Dieckman 
\& Doebeli (1999), Drossel \& Mac Kane (2000)), and find a new speciation 
equation. We present it in numerical examples, where OP can be observed.

\vspace{5mm}
\begin{center}
{\Large The concept of species orthogonalization}
\end{center}
We can look on species orthogonalization from different levels of the OP 
activity. In a self-organized molecular system (Eigen (1971)), different 
types of molecules play different roles, expressed by their microscopic 
structures. Evolution processes leading to separation of these roles can be 
called {\it orthogonalization on the micro-level} (OMIL). Since correlations 
are also built between clusters of cooperating but different types of 
molecules, natural orthogonalizing units are {\it molecular quasispecies} 
(Eigen \& Schuster (1977), Eigen, M., McCaskill, J, \& Schuster, P., (1989)).
Therefore, eigensolutions of the Eigen's equations implicitly incorporate 
quasispecies orthogonalization on the micro-level. 

The molecular system can be a part of a macro-system, competing or cooperating
with other macro-systems of the same or different types. Different types 
of macro-systems (individuals in different species or different organs inside 
individuals) analogously develop complementary functions and signaling, 
activities in different 
times or space regions.  Processes leading to such a differentiation 
of biological species can be called {\it orthogonalization on the macro-level} 
(OMAL). We could again find that OMAL is rather realized between 
self-organized clusters of species, analogously to the quasispecies on 
the micro-level, which can help to improve species definitions.
Forces leading to OMAL have predominantly macroscopic origin, but their
influence must be ``inheritably" fixed on a micro-level, which can be 
realized there as a part of ``externally driven" OMIL. 

\vspace{3mm}
\begin{center}
{\small REALIZATION OF ORTHOGONALIZATION PROCESSES}
\end{center}
Realization of OP on all structural levels of biological species is rather
fascinating. Starting from the micro-level, proteins on the tertiary level,
for example, resemble macro-tools, where prototypical sections can be 
identified with letters in a hypothetical alphabet (Lesk (1951), Creighton 
(1992), Holm \& Sander  (1998)), showing the activity of OMIL, {\it i.e.} 
a functional orthogonalization of certain protein sections. 
Recently, for example, orthogonalization as a measure 
of similarity between different molecules was used in a modeling of 
prebiotic species formation (Segr\'e {\it et al.} (2000)).  Proteins also 
reflect the environment temperature, pressure or acidity by the compactness 
of their folding (Lumsden {\it et al.} (1997)), which shows that some 
degrees of freedom can be tuned in proteins from the macro-level, 
without qualitatively affecting their {\it functions}.

Comparative studies show (Creighton (1992)) that proteins, common among 
different species, have nearly zero mutations at some regions, reflecting 
functional importance of these sections. Other sections are nearly {\it 
neutral} with respect to protein's functions, since compositions of 
amino-acids present here are reasonably varied between evolutionary distant
species (Wang (1996)). Such a structural ``self-averaging" often applies to 
elongated sections stretching out of the proteins. The fact that 
fluctuations in these sections between individuals in a single species are 
still {\it statistically small} could reveal their use for ongoing 
OMAL processes. These processes can be engaged in a structural tuning 
of the neutral sections, with the goal to influence speed (catalytic 
strength), timing and other aspects of protein's functions, and thus 
vary the species {\it morphology} 
and general activity. Some genes, present in the gene-pool in many 
similar copies (Creighton (1992)), are also tuned by OMAL to perform the 
same functions in different evolution stages of the organism, which 
carries them. Therefore, proteins and other microscopic units probably 
first internally develop by OMIL, and later are being externally 
tuned by OMAL, according to the needs on the macro-level. 

OMAL is realized through an information (vertical) flow between the 
species macro and micro-levels, 
where each level expresses its needs and possibilities in a specific way.  
Stimulations between the levels are rich: the macro-world provides 
species individuals with various resources, acts on them through a set of {\it 
macroscopic variables} $\Phi$ (like radiation, temperature, humidity, ...), 
and checks their ability to fit a given time-dependent environment. The 
individuals react according to their inherited and gained informations 
(Irwin \& Price (1999)), coded in their gene-pools and memories, evolving 
as a response to the living conditions. 

\vspace{5mm}
\begin{center}
{\Large Description of orthogonalization processes}
\end{center}
Mathematically, an individual can be represented by 
a point in a vector space, where coordinates $\phi_i$ 
are related to codons, genes, or other microscopic units (Baake 
{\it et al.} (1997)). Giving the fluctuations in their composition,
a species with many individuals occupies a finite volume $\Delta \phi$ 
of this space. An extreme example are viral quasispecies (Eigen (1993)), 
where $\Delta \phi$ is very large and rapidly grows in time by mutations.  
The species individuals can be also described in a vector space with 
coordinates given by their macroscopic characteristics ${\bf x}$ 
(Pr\"{u}gel-Bennett (1997)). Some sets of microscopic 
parameters $\phi$ (genotypes) can lead to morphological solutions 
with close macroscopic parameters ${\bf x}$ (phenotypes). 
Ambiguity of this reflection can be traced by a genotype-phenotype 
mapping (Fontana \& Schuster (1998)). 

The two types of vector spaces are formed by a direct product of their 
subspaces ${\cal H}= {\cal H}_1 \times {\cal H}_2 \times ... \times 
{\cal H}_n$ for the involved microscopic or macroscopic parameters.  
In dependence on the studied level, an individual can be thus identified 
with the vector ${\bf x} =(x_1,x_2,...,x_n)$ or $\phi=
(\phi_{1}, \phi_{2}, ..., \phi_{n})$. A species 
with many individuals can be described by a (character) distribution 
$f({\bf x})$ or $\tilde{f}(\phi)$, which reflect various correlations 
in the populations. The distributions can be smoothed and normalized 
$f_0({\bf x})=f({\bf x})/N$, where $N$ is the number of involved individuals.

\vspace{3mm}
\begin{center}
{\small EVALUATION OF ORTHOGONALIZATION PROCESSES}
\end{center}
To quantitatively appreciate OP, we can formally introduce a measure of 
species orthogonality. We say that two species are {\it orthogonal with 
a weight} $\mu$, if a scalar product (Prugove\v cki (1981)) of their 
distributions $\langle f_0| g_0 \rangle$ fulfills 
\begin{equation}
\langle f_0| g_0 \rangle \equiv \int dx_1 dx_2 ... dx_n\
\sqrt{f_0(x_1,x_2,...,x_n) \ g_0(x_1,x_2,...,x_n)} = 1-\mu\ .
\label{FG}
\end{equation}
Any species is orthogonal to itself with a weight $\mu=0$. Similarly, 
$\mu=1$ for any two species, if their distributions for at least one 
component of the vector ${\bf x}$ totally avoid each other.  Definition 
(\ref{FG}) just touches the aspect of a ``visual dissimilarity" of the 
character distributions in species. The value $\mu \approx 1$, 
for example, does not assure that the two involved species are related 
in any way, and if they are, it does not prove that OP are {\it completed}. 
The last should be independently checked from the time-independence of 
the distributions $f$, $g$, as outlined in the Introduction.

To get more refined tools for capturing OP, we should rather focus on 
some ``dynamical aspects" of OP, like the fact that species living in 
close contacts often largely {\it interact}, even if they are 
practically orthogonal $\langle f| g \rangle \approx 0$.  Interactions, 
which enable orthogonalization, are realized through competition, 
described here by broadened distributions $f_C$, $g_C$ (see Eqn.~\ref{LAND}). 
For interacting species, these functions thus have a nonzero overlap 
$\langle f_C| g_C \rangle\neq 0$, even if $\langle f| g \rangle \approx 0$. 
When the interactions proceed vertically on the food-web (Amaral \& Meyer 
(1999)), OP can be traced analogously.

It is also important to know in which way OP form {\it correlations} between 
the parameters $x_1,x_2,...,x_n$ of interacting species. This can be tested 
from the species distributions $F(x_k,x_l, ...)$, $G(x_k,x_l, ...)$, 
resulting by projection of $f({\bf x})$, $g({\bf x})$ on the selected
$x_k,x_l, ...$ parameters
\begin{equation}
F,G(x_k,x_l, ...)=\int dx_1 dx_2 ... dx_{k-1} dx_{k+1} ... 
dx_{l-1} dx_{l+1} ... dx_n~
f,g({\bf x})\ .
\label{FPROJ}
\end{equation}
Similarly as $f_C$, $g_C$, the projected distributions $F$, $G$ could 
also overlap 
\begin{equation}
\langle F|G \rangle \equiv \int dx_k dx_l ...~ \sqrt{
F(x_k,x_l, ...) G(x_k,x_l, ...)}\neq 0\ ,
\label{FGPROJ}
\end{equation}
even if the species are quasi-orthogonal $\langle f| g \rangle \approx 
0$. For interacting species, this can be largely due to correlations
formed by OP between the individual parameters $x_{1,2,...,n}$, in contrast
to the overlap of $f_C$, $g_C$ due to broadening (interaction).

\begin{figure}[htbp]
\vspace*{-3mm}
\hspace*{25mm}
\epsfxsize=120mm
\epsffile{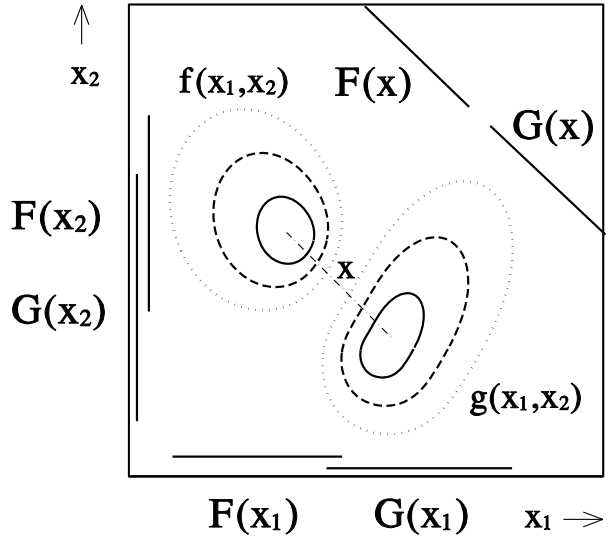}
\vspace*{-9mm}
\caption{Scheme of the distributions $f(x_1,x_2)$, $g(x_1,x_2)$ for two
species, quasi-orthogonal ($\langle f|g \rangle \approx 0$) in the
parameters $x_1$, $x_2$. The projected distributions $F(x_1)$, $G(x_1)$ or
$F(x_2)$, $G(x_2)$ overlap $\langle F(x_{1(2)})|G(x_{1(2)}) \rangle
\neq 0$, while the distributions $F(x)$, $G(x)$ resulting by the
projection along the effective orthogonalization variable $x$ are still
nearly orthogonal $\langle F(x)|G(x) \rangle \approx 0$.}
\label{SPE0}
\end{figure}

These correlations can be traced, if we project $f$, $g$ on a {\it 
functionally relevant cluster} of parameters. An example for two interacting 
bird species might be the triplet: $x_b$-beak length,  $x_l$-leg length 
and  $x_w$-wing size. This triplet could give a reasonably small overlap 
$\langle F|G \rangle \approx 0$, while projections on subsets of $(x_b,x_l,
x_w)$ might already result in large overlaps $\langle F|G \rangle \neq 0$.  
This can be especially relevant for the projected and competition 
broadened distributions $F_C$, $G_C$. We can thus define an {\it 
effective orthogonalization variable} $x$, which would give the least 
overlap when we project on it. This variable then speaks about the 
character of correlations built by OP. The simplest possibility how
to define it (see Fig.~\ref{SPE0}) is to project along the vector 
${\bf x} =({\bf C_F}-{\bf C_G})\, x$, oriented in the direction between 
the centers of gravity ${\bf C_F}=(x_b^F, x_l^F, x_w^F)$ and  ${\bf C_G}
=(x_b^G, x_l^G, x_w^G)$ of the projected species distributions
$F(x_b,x_l,x_w)$, $G(x_b,x_l,x_w)$, and integrate these distributions 
out in the remaining two orthogonal directions. We will apply the 
effective one dimensional variable $x$ in our numerical studies.

In Fig.~\ref{SPE0} we schematically present the distributions $f(x_1,x_2)$, 
$g(x_1,x_2)$ for two species. They are practically orthogonal 
$\langle f(x_1,x_2)| g(x_1,x_2)\rangle \approx 0$, while the projected 
distributions $F(x_1)$, $G(x_1)$ or $F(x_2)$, $G(x_2)$ are not 
$\langle F(x_{1,2})|G(x_{1,2}) \rangle \neq 0$.  When we project 
$f(x_1,x_2)$, $g(x_1,x_2)$ on the effective orthogonalization variable 
$x$, defined by the direction between the distribution ``centers", the 
resulting projected distributions $F(x)$, $G(x)$ become also nearly 
orthogonal $\langle F(x)|G(x) \rangle \approx 0$. This reveals the
OP-built correlations between parameters $x_1$, $x_2$ in the two species.

\vspace{7mm}
\begin{center}
{\Large Orthogonalization in sympatric speciation}
\end{center}
Investigation of OP in species is a demanding task, due to the formation of 
the complex correlations between species parameters and the different 
involved levels. OMAL processes can be often observed in evolutionary or 
functionally close species. Very interesting from the point of OP is the 
{\it sympatric speciation}, with its many mutually interconnected aspects, 
like the presence of competition or the role of assortativity in sexual 
interbreeding (Doebeli (1996b), Kondrashov \& Kondrashov (1999), Dieckman 
\& Doebeli (1999)). The activity of OMAL could help us to understand some 
problems of sympatry, like its possible existence on a ``flat landscape" solely 
due to competition (Rosenzweig (1978)). 

\vspace{3mm}
\begin{center}
{\small SPECIATION EQUATION}
\end{center}
Here, we model sympatric speciation on the macro-level, where we investigate 
the presence of OMAL. For simplicity, we do not explicitly model here
the parallel 
fixation on the micro-level. Only sexually reproducing organisms are 
considered, but the results could be extended also to clonal species.  We 
describe the total character distribution $f({\bf x})$ by the new equation 
\begin{eqnarray}
\frac{\partial f({\bf x})}{\partial t} = {\cal R}(\Phi,{\bf x}) 
\int \int d{\bf x}_I~ d{\bf x}_{II}~ {\cal S}({\bf x}_I,{\bf x}_{II})\ 
\Bigl( f({\bf x}-{\bf x}_I)~ f({\bf x}+{\bf x}_{II}) \Bigr)^{\alpha}
- \frac{f({\bf x})}{\tau({\bf x})}\ ,
\label{tran}
\end{eqnarray}
and assume that $f({\bf x})$ splits into several sub-populations $f_i({\bf x})$
for {\it individual species} with a limited interbreeding. The front term on 
the r.h.s., representing growth of the 
species, is controlled by the effective resources ${\cal R}$, with abiotic 
and biotic components, and dependence on factors from the set $\Phi$, like 
temperature or radiation. It is also influenced by the sexual function 
${\cal S}$ with a second order mating (two partners) in a power $\alpha$. 
The last term in Eqn.(\ref{tran}) represents dying of individuals with a 
time $\tau({\bf x})$, due to natural reasons, predator-prey coupling or 
catastrophes. It can also reflect possible extinctions, since $f({\bf x}$) 
would keep a constant norm for $\tau({\bf x})\approx \infty$. 

\vspace{3mm}
\begin{center}
{\small RESOURCES FUNCTION}
\end{center}
The resources function ${\cal R}$ plays the role of a ``fitness landscape",
considered in other speciation studies (Peliti (2000)). We use it in the form
\begin{eqnarray}
{\cal R}(\Phi,{\bf x}) = {\cal R}_0(\Phi,{\bf x})\,
\exp\Bigl\{ - \Bigl( \int d\bar{{\bf x}}~
 {\cal R}^C(\Phi,{\bf x}-\bar{{\bf x}})
f(\bar{{\bf x}}) \Bigr)^{\beta} \Bigr\}\ ,
\label{FUN}
\end{eqnarray}
where competition can be largely varied. Here, ${\cal R}_0 (\Phi,{\bf x})$ 
represents all the resources available for an 
individual with parameter ${\bf x}$, in the absence of other individuals, 
and the external parameters $\Phi$. We simply assume that it is given by the 
Gaussian function ${\cal R}_0({\bf x})=R_0~ e^{ -|{\bf x}-{\bf x}_R|^2/
(\sigma_R)^2 }$. Exploitation of the resources is controlled by the 
exponential function in Eqn.(\ref{FUN}) with the distribution 
$f({\bf x})$, which is competition-broadened by the function ${\cal R}^C$.
This broadening expresses the fact that individuals with parameter 
${\bf x}^{'}$ can eat the food already consumed by individuals with 
parameter ${\bf x}$. If we assume that ${\cal R}^C$ is also Gaussian, 
its width $\sigma_C$ determines the effective ``distance" of the 
consumption/competition, and additional nonlinearity of this process 
can be tuned by the parameter $\beta$. 

Roughening of the landscape ${\cal R}_0(\Phi,{\bf x})$ reflects the discrete 
character of species, which form the resources, and the structured ability 
of other species to consume them. In principle, both can be related to a 
``limited scaling" in the species parameters. For example, individuals 
of some species cannot get just smaller without qualitatively changing 
their properties, like the ability to hide by digging a hole, because of 
the increasing danger from predators, as they become smaller. In this way, 
complex correlations in species parameters are built by OMAL, which are 
roughening the landscape ${\cal R}_0(\Phi,{\bf x})$.  We do not explicitly 
consider here the food-web structure, where landscape roughening could 
propagate, and neglect also predator-prey dynamics. 

\vspace{3mm}
\begin{center}
{\small SEXUAL TERM}
\end{center}
In the growth term from Eqn.(\ref{tran}), we assume that the progeny population 
$f({\bf x})$ receives the properties ${\bf x}$ from the parents with 
distributions $f({\bf x}-{\bf x}_I)$, $f({\bf x}+ {\bf x}_{II})$. In the 
{\it mean heritage} approximation, only parents with ${\bf x}_I={\bf x}_{II}$ 
contribute to this solution. The dependence of the birth rate on the distance of
parents parameters ${\bf X}={\bf x}_I +{\bf x}_{II}$ is given by the sexual 
correlation function ${\cal S}({\bf X}, {\bf \xi})$. Declination of the 
progeny properties from this mean solution is described by the difference 
$\xi={\bf x}_I -{\bf x}_{II}$, which originates in mutations and other 
effects, like gene mixing and activation. We can again assume that 
${\cal S}$ has a Gaussian form 
\begin{equation}
{\cal S}({\bf X},\xi)=S^0\ \frac{e^{-|{\bf X}|^2/\sigma_{S}^2
-|\xi|^2/\sigma_{M}^2 }}{2\pi \sigma_{S} \sigma_{M}}\ ,
\label{CS}
\end{equation}
where the width of sexual interests $\sigma_{S}$ and the declination 
$\sigma_{M}$ can be partly tuned by OMAL.  The factor $\alpha$ in 
Eqn.(\ref{tran}) reflects the ``power of interbreeding".  Speciation is 
rather sensitive to the departure of the equation from the quasi-linearity 
$\alpha \approx 0.5$, as we discuss below.  The nonlinearity could be 
also incorporated 
in a more general version of the decay term $-f({\bf x})/\tau({\bf x})$ in 
Eqn.(\ref{tran}), where it would reflect a decreased defense ability of 
small size populations.

\vspace{5mm}
\begin{center}
{\Large Numerical Studies}
\end{center}
In numerical studies, we use projected distributions with just one scalar 
property $x$ for the {\it effective orthogonalization variable} (see 
Fig.~\ref{SPE0}), in which the species 
overlap the least. We assume, for simplicity, that the projected 
distribution $F(x)$ (called here also $f(x)$) follows the same 
Eqn.~\ref{tran} as the full distribution $f(x_1,x_2,...,x_n)$. We define 
the formed species $i$ as separate sub-populations $f_i(x)$ in the character 
distribution $f(x)$, similarly as in (Doebeli (1996b), Drossel \& Mac Kane 
(2000)). The species $i$, $j$ are considered to be orthogonal if their 
final steady-state distributions $f_i(x)$, $f_j(x)$ do not overlap 
$\langle f_i|f_j \rangle \approx 0$. The sub-populations of the 
consumption/competition function $f_C(x)$ can still overlap 
$\langle f_{Ci}| f_{Cj} \rangle \neq 0$, which leads to the fact that 
the species separation in the variable $x$ is mostly controlled by 
the finite consumption/competition width $\sigma_C$. 

We transform in Eqn.~\ref{tran} the internal variables ($x_I$, $x_{II}$) 
or equivalently ($X$, $\xi$) to the new variables ($z$, $\bar{z}$), where 
$X=2\bar{z}$, $\xi=2(x-z)$. Then the equation reads 
\begin{eqnarray}
\frac{\partial f(x)}{\partial t} = {\cal R}(x)~ \int \int 8~ dz~d\bar{z}~ 
\frac{e^{-4\bar{z}^2/\sigma_S^2-4(x-z)^2/\sigma_M^2 }}{10\pi\sigma_S\sigma_M}~
\Bigl( f(z-\bar{z})~ f(z+\bar{z}) \Bigr)^{\alpha}-\frac{f(x)}{\tau}\ ,
\label{SPEC}
\end{eqnarray}
where, the resources function is
\begin{equation}
{\cal R}(x)=R_0~ e^{-x^2/\sigma_R^2-a_C\, \bigl( f_C(x) \bigr)^\beta }\ , \ \ 
f_C(x) = \int d\bar{x}~ \frac{ e^{-(x-\bar{x})^2/\sigma_C} }
{ \sqrt{2\pi}\sigma_C } f(\bar{x})\ .
\label{LAND}
\end{equation}
In Eqns.(\ref{SPEC}-\ref{LAND}) we use the values $S^0=0.2$ and $a_C=4$
to easily present numerical results for different functions. 

\vspace{3mm}
\begin{center}
{\small PRACTICAL EXAMPLES}
\end{center}
We use Eqns.(\ref{SPEC}-\ref{LAND}) to study under which conditions 
sympatric speciation can be obtained on broad Gaussian resources, and
pay also attention to the role of their local inhomogeneities.
We also specify the role of OMAL in the obtained solutions.

\begin{figure}[htbp]
\vspace*{-15mm}
\hspace*{20mm}
\epsfxsize=310mm
\epsffile{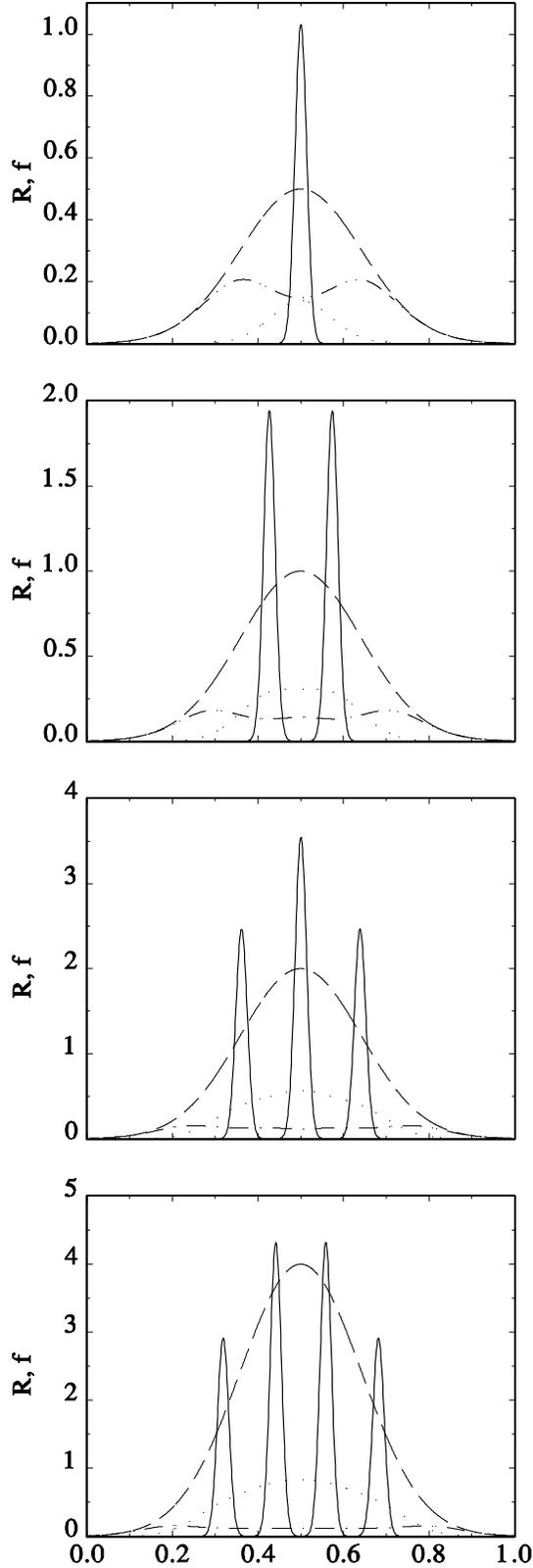}
\vspace*{-15mm}
\caption{Species population $f$ (competitive population $f_C$) in a sympatric
speciation is shown by full (dotted) lines as a function of the effective
parameter $x$. The dashed (dash-dotted) lines represent the free (competitive)
landscape of resources. The plots with one to four species correspond to
strength of resources $R_0=0.5$, $1$, $2$, $4$ and other parameters in
the text.}
\label{spe1}
\end{figure}

In Fig.\ref{spe1} we present the steady-state (globally stable) numerical
solutions of Eqns.(\ref{SPEC}-\ref{LAND}) for species formed by sympatric 
speciation.  The dashed and dash-dotted curves correspond to the free and 
unused resources without and with the competition exponential from 
(\ref{LAND}), respectively.  The solid and dotted curves represent the 
{\it total} population $f(x)$ and the consumption/competition function 
$f_C(x)$, respectively. The situations with one to four species are obtained 
for the Gaussian 
resources of the width $\sigma_R =0.2$ and strengths R$_0=0.5$, $1$, $2$, $4$. 
The consumption/competition width is $\sigma_C=0.1$, the sexual and mutation 
widths are chosen the same $\sigma_S= \sigma_M= 0.015$, the power parameters 
are $\alpha= \beta=0.6$ and the lifetime is $\tau=100$. 

The broad Gaussian 
resources can be seen as ``quasi-flat", since $\sigma_R$ is larger than 
other parameters in the model. When the richness of resources R$_0$ is 
rather small, only $\sigma_C$ is close in size to $\sigma_R$. As R$_0$ 
increases, an {\it effective width} of resources $\Sigma_R$ (on which 
species can survive) becomes larger than $\sigma_R$, so the resources 
can feed more and more species.  If $\sigma_R$ alone (not R$_0$) is enlarged 
to get a nearly flat landscape, the number of species grows, but their 
separation is always given by $\sigma_C$. If $\sigma_C$ does not vary much,
the resulting self-organized {\it speciation quasi-periodicity} as a 
function of $x$ determines a universal {\it speciation volume}.  In the 
present broad Gaussian landscape, with an exponential form of 
competition (see (\ref{LAND})), speciation can be obtained only if the 
power of interbreeding is slightly nonlinear $\alpha>0.5$. This adds
to the speciation conditions on a quasi-flat landscape.

The unused resources $( - . - . )$ in Fig.\ref{spe1} are {\it flattened} 
by the total distribution $f_C(x)$, which largely copies the shape of the
resources  $( - - - \ - )$, even though the total distribution 
$f(x)$ has sharply separated sub-populations. This shows an efficient 
sharing of resources by the orthogonalized consumption needs of 
the species (Roughgarden (1976)). We have used Eqn.~\ref{FGPROJ} to
calculate the relative overlaps $r_{Cij}=\langle f_{Ci}|f_{Cj} 
\rangle/\langle f_C|f_C \rangle$ of the individual contributions $f_{Ci}(x)$ 
to $f_C(x)$. These $f_{Ci}(x)$ can be obtained from Eqn.~\ref{LAND},
where we substitute the individual distributions $f_i(x)$.  The situations 
with $n=2-4$ species in Fig.\ref{spe1} give: $n=2$, $r_{12}=0.296$; $n=3$, 
$r_{12}=0.219$, $r_{13}=0.045$ and  $n=4$, $r_{12}=0.17$, $r_{13}=0.062$, 
$r_{14}=0.008$, $r_{23}=0.213$.  The values $r_{Cij}$ show the steady-state 
strength of coupling of the orthogonal species ($\langle f_i|f_j \rangle 
\approx 0$). The related pressures can be {\it relaxed by inhomogeneities} 
in the broad Gaussian resources. 

Note also that the final distributions $f_i(x)$ in Fig.\ref{spe1} do not 
depend on the size of the initial Gaussian population used in iterations of
Eqns.(\ref{SPEC}-\ref{LAND}), unless this is very small. Then, for the 
present parameters, initial populations which are by 2-3 orders smaller 
than the final population tend to follow the single-peaked landscape. 
This means that the final solution gives $2n-1$ species where $2n$ species 
would normally appear, so the initial condition is ``frozen in" the final 
state of a reasonable stability. For even smaller initial populations 
species die out, due to finite lifetime $\tau$. 

Stepwise speciation resembling our results has been observed, 
for example, in stickleback species in small lakes of the coastal British 
Columbia (Schluter \& Mac Phail (1992)). Speciation driven by the increased 
strength of resources could be also the reason of diversity gradients of 
marine life, as induced by the solar radiation in coastal regions 
(Roy {\it et al.} (1998)). Similarly, richness gradients of other species 
observed towards the equator (Rapoport (1975)) could be the consequence
of {\it enlarged radiation, temperature and food zones}.

\begin{figure}[htbp]
\vspace*{-8mm}
\hspace*{20mm}
\epsfxsize=330mm
\epsffile{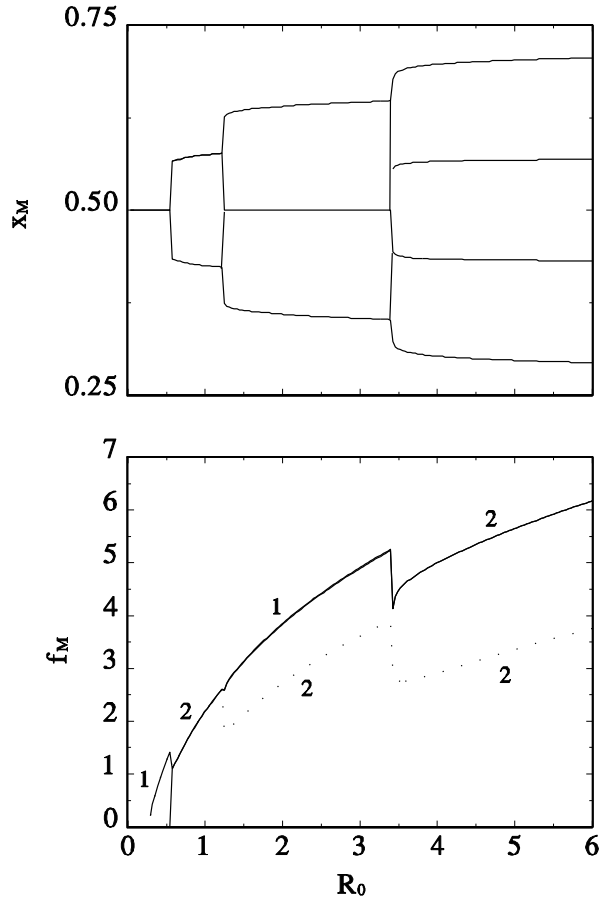}
\vspace*{-125mm}
\caption{(top plot) The maxima $x_M$ of the populations, for the
species with parameters in Fig.~\ref{spe1}, as a function of the resources
strength $R_0$. (bottom plot) The same dependence for the maxima
of the species populations $f_M$, with the number of peaks denoted.}
\label{spe2}
\end{figure}

In Fig.\ref{spe2} we show the size $f_M$ and position $x_M$ of the 
sub-population maxima as a function of the resources strength $R_0$ and 
other parameters as in Fig.~\ref{spe1}. In the top plot for $x_M$, we 
can see that after a new 
species emerges the positions $x_M$ for the peaks slightly move. Since the 
Gaussian resources rapidly fall down at the tails, the effective width 
$\Sigma_R$ increases very slowly with $R_0$, and so does the number 
of species $n$ for $n>3$.  For example, the fifth species appears only at 
$R_0\approx 13$. In the bottom plot, we show the dependence of the maxima 
$f_M$ on $R_0$, and add the number of peaks in each curve. As $R_0$ 
grows and the number of species increases from odd to even, the main maxima 
jump in values for this shape of resources, while in transitions
from even to odd the changes are reasonably continuous.  The separation of 
species parameters (character shift) shows that OMAL is accompanied by 
a {\it pressure} between the coexisting species (Slatkin (1980)), induced 
by the overlap of their $f_{Ci}(x)$ distributions. The activity is 
especially large close to speciation events, where the species parameters 
move. Experimentally observed pressure induced variation of the character 
shift, as the species move between an island and a continent, is discussed 
for example by Schluter (1988).  

\begin{figure}[htbp]
\vspace*{-8mm}
\hspace*{20mm}
\epsfxsize=330mm
\epsffile{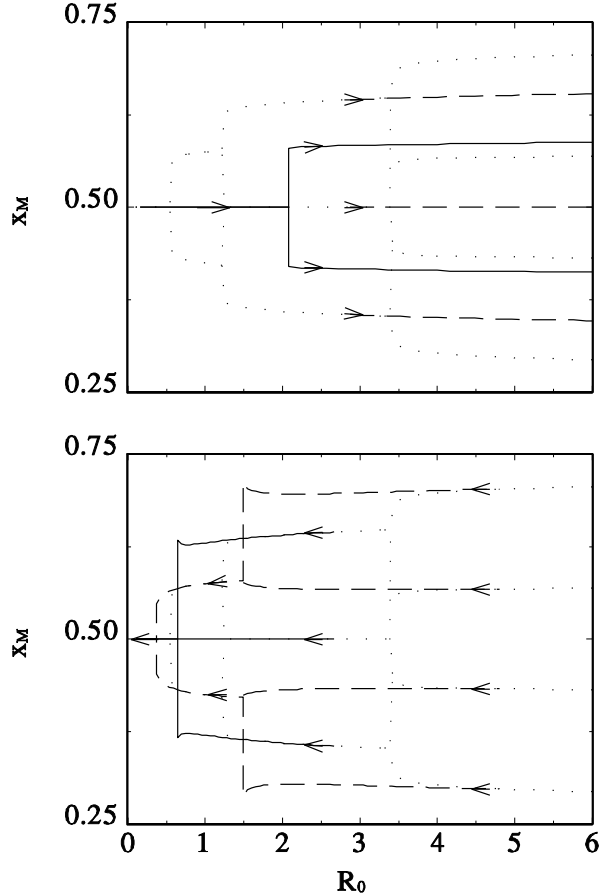}
\vspace*{-125mm}
\caption{Positions $x_M$ of the species as the resources richness R$_0$ is
slowly varied. The upper (lower) plot shows situations for increasing
(decreasing) R$_0$. Thin doted lines correspond to the globally stable
solutions from
Fig.\ref{spe2}. In the upper plot thick solid (dashed) lines present
runs starting from regions with a single (triple) species. In the lower
plot the solid (dashed) lines present runs starting from regions with
a triple (quadruple) species. In all cases speciation freezing, dependent
on history, can be observed.}
\label{spe2n}
\end{figure}

Population branching could be also studied in a time region (Kisdi (1999)). 
In our work, globally stable solutions in Fig.\ref{spe2} are calculated 
by using at each point R$_0$ a Gaussian initial population comparable 
in size with the final value. We will also study a time evolution, 
but limit to the investigation of speciation 
for an {\it adiabatic} time evolution of the resources richness R$_0$; we 
slightly change R$_0$ in each time run, and start from the populations of the 
previous R$_0$. The results for the species positions $x_M$ are shown 
in Fig.\ref{spe2n}, where thin doted lines correspond to the solutions 
from Fig.\ref{spe2}. In the upper plot, we show by thick solid (dashed) 
lines runs with 
increasing R$_0$ starting from the situation in the single-species 
(triple-species) region. In the former case, the first speciation occurs 
delayed in the double-species region, while three species would appear only
deeply in the region where four species should be; a start from the 
double-species region follows this solution as well. If we start from 
the triple-species region, the number of species keeps unchanged for the 
used parameters. 
If the resources become poorer, the situation is rather different, as 
we show on the bottom plot. A start from the four-species region gives
evolution skipping over solutions with three species, but it ``smoothly 
transfers in advance" to 
a solution with two species, and the solution with one species appears
delayed. Starting from the triple-species region, the evolution skips 
over solutions with two species, and later jumps to a solution with one 
species. Thus, if R$_0$ increases, all numbers of species appear, but
they are delayed, while if R$_0$ decreases, sequences with odd or even 
number of species seem to be realized. Since the frozen nonequilibrium 
solutions are locally stable with respect to perturbations, the system 
experiences {\it hysteresis} in the number of species if R$_0$ slowly 
oscillates with a large enough amplitude.  It is likely that the species 
parameters would change by OMAL, to partially follow the slow evolution 
of resources and bypass the mounting pressures from a fixed number of 
species. The crucial aspect is the ratio of timescales of these effects.
In nature, such a behavior can be also largely stabilized by year seasons 
and other random effects, but if we take into account the evolution on 
micro-level one could essentially face only {\it irreversible behavior}. 

\begin{figure}[htbp]
\vspace*{-15mm}
\hspace*{20mm}
\epsfxsize=305mm
\epsffile{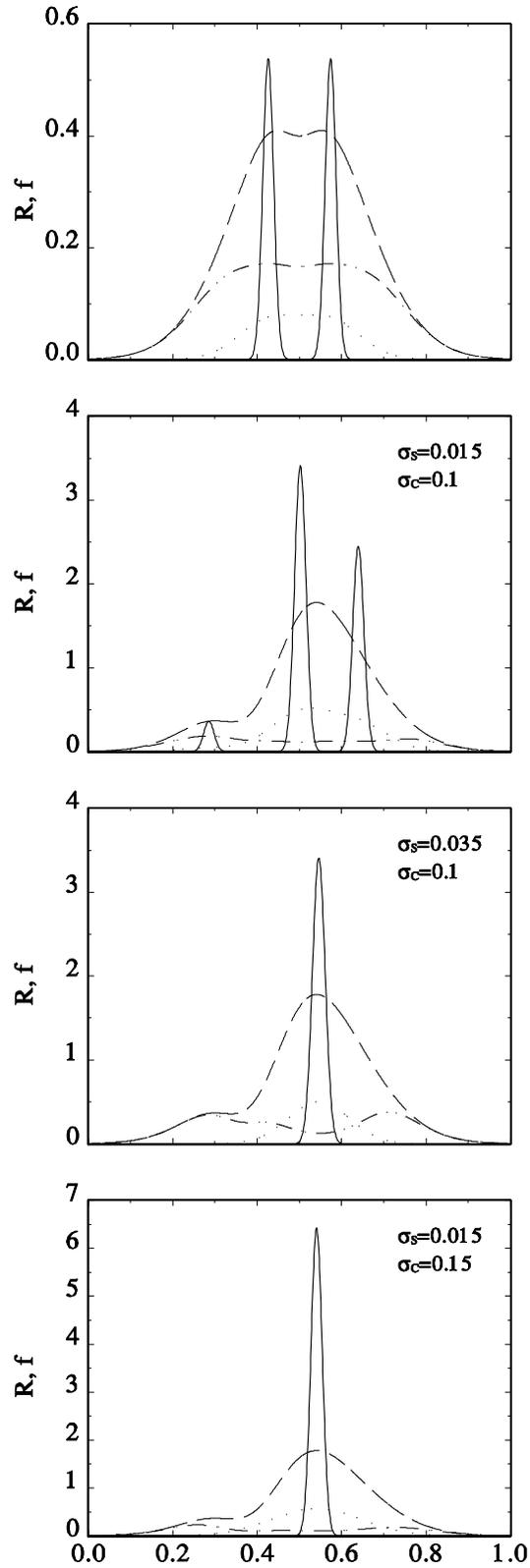}
\vspace*{-15mm}
\caption{The situations from  Fig.\ref{spe1} presented for a perturbed
Gaussian landscape. The top plot shows how the single species from
Fig.\ref{spe1} breaks in two, when the resources are flattened. Next,
one of the side species, from the third plot in Fig.\ref{spe1}, shifts and
decreases in size to fit the new landscape.
In the third plot, for the same parameters, but the sexuality width
$\sigma_S=0.035$, the side peaks dissapear, but the resources are not well
explored. This is improved on the last plot, where, instead of
$\sigma_S$, the consumption width is broadened to the value $\sigma_C=0.15$.}
\label{spe3}
\end{figure}

In Fig.\ref{spe3} we show speciation in the previous landscapes, modified 
by additional disturbances. The disturbances can {\it trigger speciation} 
at smaller competition pressures, induced by the $f_{Ci}$ distributions; 
speciation could even take place for $\alpha=0.5$, which would not 
be possible on the broad Gaussian landscape.  In the upper plot, we take 
the situation with one species in Fig.\ref{spe1}, and subtract from the 
resources $R_0~ e^{-x^2/\sigma_R^2}$ the exponential $R_1~ e^{-x^2/\sigma_F^2}$ 
with $R_1=0.1$, $\sigma_F=0.07$. The resulting landscape flattening
can split the single species in two sub-populations, since the resources 
become more efficiently explored by two species.  We have also found 
such splitting/joining jumps for the case of two or three species. 
For $R_0=1.5$ the number of species goes from three to two, while for 
$R_0=2.5$ it goes from three to four, copying thereby in both cases the 
flatter shape of resources. 

From a theoretical point of view, we could also consider that $x$ represents 
one of the parameters $\Phi$, like temperature, radiation or humidity. Then, 
the resources show how much food for the species is available at this 
parameter $x$, and other changes are analogous. Therefore, 
we should in principle be able to describe certain situations with 
{\it allopatric speciation}. For example, splitting of a Drosophila 
population, similar to the top part of Fig.\ref{spe3}, was observed 
on a microsite in a valley with two differently irradiated slopes, giving a 
radiation/temperature space gradient corresponding to our parameter $x$
(Harry {\it et al.} (1999)). A sharply different character of the slope 
sides could give flatter resources in the middle radiation/temperatures 
with side maxima as in Fig.\ref{spe3}. 

In the second plot of Fig.\ref{spe3}, we present the population for the 
situation with three species in Fig.\ref{spe1}, where we subtract, similarly 
as before, the shifted exponential $R_1~ e^{-(x+x_1)^2/\sigma_F^2}$ with 
$R_1=1$, $x_1=0.1$ and $\sigma_F=0.07$. Due to this asymmetry of the resources, 
the first side species largely shifts and decreases, in order to efficiently 
explore the new landscape,  but the total number of species still remains 
the same. This inhomogeneity-shifted species competes a little with the 
other species.  In the next plot, we use the same resources, 
but we broaden the sexual width to $\sigma_S=0.035$. This 
decreased assortativity fully prevents speciation, and the resources are 
poorly used in the tails. Much better exploration is obtained in the 
last plot, where, instead, the consumption width is increased by $50$ \% to 
$\sigma_C=0.15$, so the single species population is also larger. 
The width $\sigma_M$ is also crucial, since it can broaden and smooth 
the population.  For example, the single peaked distribution in the third 
plot of Fig.~\ref{spe3} can be also obtained if $\sigma_M$ is increased 
to the value $0.035$. 

\vspace{3mm}
\begin{center}
{\small GENERAL TUNING OF PARAMETERS}
\end{center}
Other parameters in Eqns.(\ref{SPEC}-\ref{LAND}) could also interplay in a
sensitive way. For example, the power $\alpha$ can fluctuate around the 
value $\alpha \approx 0.5$ to promote/suppress speciation.  The power 
$\beta$ can reflect the way of food consumption as well as possible 
deterrence actions aimed in protecting it (screening). For $\beta<0.5$ 
($\beta>0.5$) side populations become suppressed (enhanced) with respect 
to the middle populations. The absolute strength of consumption/competition 
$a_C$ can be tuned to control the size of the population. Finally, the 
time $\tau$ maintains the speciation pressure and keeps the population
stability (Smith \& Brown (1986)). The term with $\tau$ in Eqn.(\ref{tran})
diminishes individuals present in regions between the landscape maxima. 
It also gives sharper and a more regular speciation. Without 
competition and with large $\tau$, the population would simply follow 
the landscape, while with competition, it would not speciate sharply 
until $\tau$ is decreased. 

Recent studies point to the increased role of food-web resources in species 
survival (Amaral \& Meyer (1999)). Proportionality between the number of 
species and the resources richness, obtained above, can be also supported in 
different ways.  For example, in poor resources, generalist species might 
be triggered by broadening $\sigma_C$ and $\sigma_S$. Clusters of species 
can be extinct, and surviving species can broaden their distributions, so that
correlations formed by OMAL become decreased. Since mutations are less 
suppressed, speciation could eventually be also promoted in these conditions. 
If the resources get richer, individualists could do better. Species can narrow 
their consumption needs $\sigma_C$, but they need to live efficiently in 
other ways. Consequently, they also narrow their sexual interests 
$\sigma_S$ and trigger further speciation.  Since the populations 
largely copy resources, their richness does not make life easier for 
the numerous individuals. This drives the lasting self-organized evolution. 
OMAL can tune all these parameters in a limited amount to balance coexistence 
of species with an efficiency exploration of resources. The timescale 
of this activity goes over many generations, when these changes get 
fixed on the micro-level.

\vspace{3mm}
\begin{center}
{\small FIXATION OF OMAL }
\end{center}
We noted that neutral sections of proteins, with a relatively stable 
composition in each individual species, might be used for transcription
of OMAL. This can support selectionist views on the problem of {\it molecular 
evolution clock} (Ohta (1996)), which underlines the quasi-deterministic 
origin
of observed constant mutation rates in neutral sections of proteins. Recent 
observations show, that when bacteria are placed in a new environment, 
some of their genes start to mutate faster (Sniegowski {\it et al.} (1997), 
Schmid \& Tautz (1997)). Growth of RNA in vitro also shows that under strong 
selections the molecules go to the same limit form (Strunk \& Ederhof 
(1997)). These experiments point to a possibly deterministic 
role of the environment on the proteins, where OMAL takes an active role.

We can imagine, that tuning of protein catalytic strength can be regulated 
by other proteins, enforced by the complex molecular environment, evolving 
under the pressure of globally changing conditions on the macro-level. 
If these possibilities are exploited, and a larger change is needed, a 
jump in neutral positions of the tuned proteins could be induced by OMAL, 
which is analogous to avalanche ticking in other systems (Wu {\it et al.} 
(1993)). This ticking could largely go {\it in parallel} with the formation 
of new species, shown for example in Fig.~\ref{spe1}.

These ideas could be tested on bacteria, with short reproduction periods. 
A working scenario could be obtained by modifying some neutral protein 
region, and keeping the remaining gene-pool frozen by a periodical 
exchange. If OMAL is relevant, the protein would eventually evolve 
in many bacteria generations
to the same or a similar form. The problem is if OMAL can be reached 
in artificial laboratory conditions. Such experiments could also reveal 
to which extent one can speak about proteins in {\it nonequilibrium} 
with the cell environment. 

\vspace{3mm}
\begin{center}
{\Large Summary}	
\end{center}
We have introduced the concept of species orthogonalization on the micro and 
macro-level. OMAL is related with the need for formation of complementary 
functions and habits of different biological species living in close 
contacts. OMIL separates functions of molecular species, and assists 
OMAL on the micro-level. We have shortly discussed the ways of evaluating
orthogonalization processes.

As a practical example, we have investigated OMAL in sympatric speciation, 
but limited so far our study to the macro-level. To this goal, we have found 
a new population equation, and presented it in several numerical examples. 
Its solutions, for the projected distribution with one effective variable $x$,
show that the number of species grows in richer resources, divided between 
the species. This could help to explain the increase of species diversity 
close to the equator, where sun radiation provides richer resources. 
We have also observed speciation dependent on history, which might be
useful in modeling of ecology, and discussed possible ways of OMAL
fixation on the micro-level. We expect that orthogonalization mechanisms 
are active in other systems with competitive structures.

\vspace{5mm}

\noindent
\begin{center}
{\bf Acknowledgments:}
\end{center}
I would like to acknowledge T. Pavl\' {\i}\v cek for stimulating and 
encouraging discussions. I would also like to kindly thank M. Shapiro 
for a partial financial support.


\vspace{5mm}
\begin{center}
REFERENCES
\end{center}

\vspace{2mm}
\noindent
Amaral, L. A. N., \& Meyer, M., (1999). Environmental Changes, Coexistence,
and Patterns in the Fossil Record. {\it Phys. Rev. Lett.}
{\bf 82}, 652-655.

\vspace{2mm}
\noindent
Baake, E., Baake, M. \& Wagner, M., (1997). Ising Quantum Chain is
Equivalent to a Model of Biological Evolution. {\it Phys. Rev. Lett.}
{\bf 78}, 559-563.

\vspace{2mm}
\noindent
Bagnoli, F. \& Bezzi, M., (1997). Speciation as pattern formation by 
competition in a smooth fitness landscape.
 {\it Phys. Rev. Lett.} {\bf 79}, 3302-3305.

\vspace{2mm}
\noindent
Brown W. L. \& Wilson E. O., (1956). Character displacement, {\it Syst. Zool.}
{\bf 5}, 49-64.

\vspace{2mm}
\noindent
Creighton, T. E., (1992). Proteins. Structure and Molecular Properties.
({\it Freeman, New York}).

\vspace{2mm}
\noindent
Dieckman, U. \& Doebeli, M., (1999). On the origin of species by
sympatric speciation. {\it Nature} {\bf 400}, 354-357.

\vspace{2mm}
\noindent
Doebeli, M., (1996a). An explicit genetic model for ecological character 
displacement. {\it Ecology} {\bf 77}, 510-520.

\vspace{2mm}
\noindent
Doebeli, M., (1996b). A quantitative genetic model for sympatric speciation.
{\it J. Evol. Biol.} {\bf 9}, 893-909.


\vspace{2mm}
\noindent
Drossel, B. \& Mac Kane, A., (1999). Ecological Character Displacement in
Quantitative Genetic Models. {\it J. Theor. Biol.}, {\bf 196}, 363-376.

\vspace{2mm}
\noindent
Drossel, B. \& Mac Kane, A., (2000). Competitive Speciation in Quantitative
Genetic Models. {\it J. Theor. Biol.}, {\bf 204}, 467-478.

\vspace{2mm}
\noindent
Eigen, M., (1971). Selforganization of matter and the evolution of
biological macromolecules. {\it Naturwisseschaften} {\bf 58}, 465-523.

\vspace{2mm}
\noindent
Eigen, M. \& Schuster, P., (1977). The hypercycle. A principle of
natural self-organisation, Oart A: emergenece of the hypercycle.
{\it Naturwisseschaften} {\bf 64}, 541-565.

\vspace{2mm}
\noindent
Eigen, M., McCaskill, J, \& Schuster, P., (1989).  The Molecular 
Quasi-Species.  {\it Adv. Chem. Phys.} {\bf 75}, 149-263.

\vspace{2mm}
\noindent
Eigen, M., (1993). Viral Quasi-species. {\it Sci. Am.}, {\bf 269}, 42-49.

\vspace{2mm}
\noindent
Fontana, W. \& Schuster, P., (1998). Shaping Space: the Possible and the
Attainable in RNA Genotype-phenotype Mapping. {\it J. Theor. Biol.}, {\bf 194},
491-515.

\vspace{2mm}
\noindent
Harry, M., E., Rashkovetsky, T., Pavlicek, S. Baker, E.M. Derzhavets, P.
 Capy, M.L. Cariou, D. Lachaise, N. Asada and E. Nevo, (1999). Fine-scale
biodiversity of Drosophilidae in "Evolution Canyon" at the Lower Nahal
Oren Microsite, Israel. {\it Biologia} {\bf 54}, 685-705.


\vspace{2mm}
\noindent
Holm, L. \& Sander, C., (1998). Dictionary of recurrent domains in 
protein structures. {\it Proteins} {\bf 33}, 88-96.

\vspace{2mm}
\noindent
Hutchinson, G. E., (1968). When are species necessary? In R. C. Lewontin (Ed.).
{\it Population Biology and Evolution}: 177-186. Syracuse, New York: Syracuse
University Press.

\vspace{2mm}
\noindent
Irwin, D. E. \& Price, T., (1999). Sexual imprinting, learning and 
speciation. {\it Heredity} {\bf 82}, 347-354.

\vspace{2mm}
\noindent
Kisdi, E., (1999). Evolutionary Branching under Asymmetric Competition.
{\it J. Theor. Biol.}, {\bf 197}, 149-162.

\vspace{2mm}
\noindent
Kondrashov, A. S. \& Kondrashov, F. A.,  (1999). Interactions among 
quantitative traits in the course of sympatric speciation.
sympatric speciation, {\it Nature} {\bf 400}, 351-354.

\vspace{2mm}
\noindent
Lesk, A. M., (1951). Protein Architecture, {\it Oxford University Press
1991, (Ed. D. Rickwood \& B. D. Hames)}.

\vspace{2mm}
\noindent
Lumsden, Ch. J., Brandts, W. A. \& Trainor, L. E. H., (1997). {\it Physical 
Theory in Biology}, (World Scientific); (L. Luo p.380)

\vspace{2mm}
\noindent
May, R. M. \& Mac Arthur, R. H., (1972). Niche Overlap as a Function
of Environmental Variability. {\it Proc. Nat. Acad. Sci.}, {\bf 69},
1109-1103.

\vspace{2mm}
\noindent
Mayr, E., (1942). Systematics and the Origin of Species. (New York, 
Columbia University Press).


\vspace{2mm}
\noindent
Mayr, E., (1963). Animal Species and Evolution. (Belknapp Press, 
Cambridge, MA).

\vspace{2mm}
\noindent
Ohta, T., (1996). The neutralist-selectionist debate. {\it BioEssays},
{\bf 18}, 673-683.

\vspace{2mm}
\noindent
Peliti, L., (2000). Fitness Landscapes and Evolution. (cond-mat/9505003).

\vspace{2mm}
\noindent
Plotkin J. B., Potts M. D., Yu D. W., Bunyavejchewin S.,  Condit R.,
Foster R., Hubbell S., LaFrankie J., Manokaran N., Lee H.-S., Sukumar R.,
Nowak M. A., and Ashton P. S. (2000). Predicting species diversity in 
tropical forests. {\it Proc. Natl. Acad. Sci. USA}, {\bf 97}, 10850-10854.

\vspace{2mm}
\noindent
Pr\"{u}gel-Bennett, A., (1997). Modeling Evolving Populations.
{\it J. Theor. Biol}, {\bf 185}, 81-95. 
{\bf 82}, 1983-1986.

\vspace{2mm}
\noindent
Prugove\v cki, E., (1981). Quantum Mechanics in Hilbert Space. 
(NY, {\it Academic Press}). 

\vspace{2mm}
\noindent
Rapoport, E. H., (1975). Areografia: estrategias de las especieses.
{\it Mexico City: Fondo de Cultura Economica}.

\vspace{2mm}
\noindent
Roughgarden, J., (1976). Resources Partitioning Among Competing Species -
A Coevolutionary Approach. {\it Theor. Popul. Biol}, {\bf 9}, 388-424.

\vspace{2mm}
\noindent
Rosenzweig, M. L., (1978). Competitive speciation. {\it Biol. J. Lin. Soc.},
{\bf 10}, 275-289.

\vspace{2mm}
\noindent
Rosenzweig, M. L., (1995). Species Diversity in Space and Time.
({\it Cambridge University Press}, NY).

\vspace{2mm}
\noindent
Roy, K., Jablonski, D., Valentine, J. W. \& Rosenberg, G., (1998).
Marine latitudinal diversity gradients: Tests of causal hypotheses.
{\it P. Natl. Acad. Sci. USA} {\bf 95}, 3699-3702. 

\vspace{2mm}
\noindent
Schluter, D., (1988). Charactery Displacement and the Adaptive Divergence
of Finches on Islands and Continents. {\it Amer. Natur.}, {\bf 131}, 799-824.

\vspace{2mm}
\noindent
Schluter, D. \& Mac Phail, J. P., (1992). Ecological Charactery
Displacement in Sticklebads. {\it Amer. Natur.}, {\bf 140}, 85-108.

\vspace{2mm}
\noindent
Schmid, K. J. \& Tautz, D., (1997). A screen for fast evolving genes 
from Drosophila. {\it P. Natl. Acad. Sci. USA} {\bf 94}, 9746-9750. 

\vspace{2mm}
\noindent
Segr\'e, D., Ben-Eli, D. \& Lancet, D., (2000). Compositional genomes: 
Prebiotic information transfer in mutually catalytic noncovalent assemblies.
{\it P. Natl. Acad. Sci. USA} {\bf 97}, 4112-4117.

\vspace{2mm}
\noindent
Slatkin, M., (1980). Ecological character displacement. {\it Ecology}
{\bf 61}, 163-177.
\vspace{2mm}
\noindent
Smith, J. M., (1966). Sympatric Speciation.  {\it Amer. Natur.}, 
{\bf 100}, 637-650.

\vspace{2mm}
\noindent
Smith, J. M. \& Brown, R. L. W., (1986). Competition and Body Size.
{\it Theor. Popul. Biol.} {\bf 30}, 166-179.

\vspace{2mm}
\noindent
Sniegowski, P. D., Gerrish, P. J. \& Lenski, R. E., (1997). 
Evolution of high mutation rates in experimental populations of {\it E. coli}.
{\it Nature} {\bf 387}, 703-705.

\vspace{2mm}
\noindent
Strunk, G. \& Ederhof, T., (1997). Machines for automated evolution 
experiments in vitro based on the serial-transfer concept.
{\it Biophys. Chem.} {\it 66}, 193-202.

\vspace{2mm}
\noindent
Valentine, J. W. \& Walker, T. D., (1988). Diversity Trends within 
a Model Taxonomic Hierarchy. {\it Physica} {\bf 22}D, 31-42.

\vspace{2mm}
\noindent
Wang, Zhi-Xin, (1996). How many fold types of protein are there in nature?
{\it Proteins} {\bf 26}, 186-191.

\vspace{2mm}
\noindent
Wu X-l., Maloy K. J., Hansen A., Ammi M., \& Bideau D. (1993).  Why 
hour glasses tick. {\it Phys. Rev. Lett.} {\bf 71}, 1363-1366.  

\end{document}